# Néel vector rotation driven by spin-orbit torque in amorphous ferrimagnetic GdCo

Tetsuma Mandokoro[1], Yoichi Shiota[1,2*], Tomoya Ito[1], Hiroki Matsumoto[1], Hideki Narita[1], Ryusuke Hisatomi[1,2], Shutaro Karube[1,2], and Teruo Ono[1,2*]

[1]*Institute for Chemical Research, Kyoto University, Uji, Kyoto 611-0011, Japan*

[2]*Center for Spintronics Research Network, Kyoto University, Uji, Kyoto 611-0011, Japan*

Authors to whom correspondence should be addressed: Yoichi Shiota, shiota-y@scl.kyoto-u.ac.jp; Teruo Ono, ono@scl.kyoto-u.ac.jp

Spin superfluidity, a phenomenon enabling low dissipative spin transport analogous to superfluidity in liquid helium and superconductivity in electronic systems, has remained a theoretical concept. To realize the spin superfluidity in an antiferromagnet, it is necessary to excite a Néel vector rotation within the magnetic easy-plane, which has been elusive so far. In this study, we demonstrated spin-orbit torque-driven Néel vector rotation in amorphous ferrimagnetic GdCo. A pseudo-magnetic easy-plane is formed in a spin-flop state under an external magnetic field at the vicinity of the magnetization compensation temperature, and we observed stochastic binary switching in the anomalous Hall resistance, directly attributed to Néel vector rotation. Furthermore, homodyne detection confirmed rotation frequencies in the GHz range as expected from atomic spin simulations, providing evidence of dynamic Néel vector behavior. These findings represent a crucial advance towards the realization of spin superfluidity.

Efficient spin transport with minimal energy dissipation has become a focus in spintronics due to its potential to revolutionize information technologies. Compared to spin currents mediated by conduction electrons or spin waves, "spin superfluidity"—a quantum phenomenon characterized by low-dissipative spin current—stands out as a promising candidate[1,2]. Analogous to superfluidity in liquid helium and superconductivity in electronic systems, the spin superfluidity offers the possibility of transmitting spin angular momentum over macroscopic distance without significant loss[1,2].

In magnetic systems, the spin superfluidity is achieved by maintaining a rotation of magnetization through a steady injection of angular momentum[1,3–9]. In ferromagnets, dipolar interactions dramatically affect spin transport longer than the exchange coupling length, resulting in a disruption of the spin superfluidity over long distance[5,6]. Instead, antiferromagnets are promising for the spin superfluidity because they have no net magnetization and are not affected by the long-range dipolar interactions[10-14]. The spin superfluidity in antiferromagnets requires a rotation of the Néel vector within a magnetic easy-plane while maintaining a fixed net magnetization[4,6,9], as shown in Fig. 1. When spin currents are injected into an easy-plane antiferromagnet, where the spin polarization direction is orthogonal to the magnetic easy-plane, spin injection can cause the local magnetic moments to tilt slightly out of the easy-plane. This tilt generates an internal antiferromagnetic exchange field, which drives the Néel vector to rotate within the magnetic easy-plane at frequencies adjustable by the current[3,4]. Although there are enormous reports on the control of the Néel vector by spin-orbit torque (SOT)[15-23], direct observation of the Néel vector rotation has been lacking so far due to experimental challenges, such as pinning by a crystalline magnetic anisotropy within the easy-plane.

In this study, we investigate the dynamics of the Néel vector rotation in amorphous compensated ferrimagnet GdCo, which offers several advantages as a promising platform for easy-plane antiferromagnets[24]. First, its amorphous structure eliminates the intrinsic crystalline magnetic anisotropy, reducing the threshold current required to drive the spin dynamics. Second, at the vicinity of the magnetization compensation temperature, a spin-flop state is induced under relatively low magnetic fields, forming a pseudo-magnetic easy-plane[25,26]. Third, previous studies showed that GdCo enables highly efficient spin manipulation through the SOT[27-29]. Leveraging these properties of ferrimagnetic GdCo, we prepared a trilayer structure consisting of Ta/GdCo/Pt and observed stochastic binary switching in an anomalous Hall resistance after an application of a 10-$\mu$s current pulse, directly attributed to the Néel vector rotation. Furthermore, homodyne detection, which measures the rectified Hall voltage arising from coupling between microwave currents and the Néel vector rotation, revealed GHz-range rotation frequencies. These results made a significant step toward utilizing the spin superfluidity in antiferromagnetic systems.

Thin films of Ta(5 nm)/Gd$_{0.28}$Co$_{0.72}$(3 nm)/Pt(5 nm) were deposited on a thermally oxidized Si substrate by dc magnetron sputtering. The GdCo alloy layer was formed by co-sputtering of Gd and Co targets, and the composition of GdCo was controlled by the sputtering power for each element. Ta and Pt layers prevent oxidation from the substrate and the atmosphere, respectively, and serve as spin injection sources due to large charge-to-spin conversion[30-33]. Figure 2(a) shows an optical microscope image of the device structure. The films were patterned into a Hall device with a cross shape by photolithography and Ar ion milling, followed by the deposition of SiO$_2$ with a thickness of 17.5 nm to avoid contact between the device edge and the atmosphere. The width and length of the channel of the Hall device were 4 and 10 $\mu$m, respectively, and those of the Hall probe were 4 and 40 $\mu$m. The contact electrode of Ti(5 nm) / Au(80 nm) was then patterned by using photolithography and lift-off process.

We first characterized the magnetization compensation temperature and the spin flop transition via the anomalous Hall effect (AHE) measurements using a measurement setup shown in Fig. 2(a). In the case of ferrimagnets composed of transition metals (TM) and rare earths (RE), the magnetotransport properties are mainly contributed by the TM material[26,29,34,35], which is Co in this study. Figure 2(b) shows the AHE resistance R$_{xy}$ measurements under the magnetic field ($H_{ext}$) swept along the $y$-direction. Due to a slight misalignment of the magnetic field with the $z$-direction (out-of-plane direction), square hysteresis loops were observed, indicating that the magnetic easy-axis of the GdCo alloy layer is perpendicular (see also the AHE hysteresis loops under the out-of-plane magnetic field in Supplementary Fig. S1). The hysteresis loops reverse polarity at the magnetization compensation temperature $T_M$, which was determined to be 260 K. After the dominant sublattice magnetization saturated in the magnetic field direction, the bending of the AHE resistance was observed at specific magnetic fields (arrows in Fig. 2(b)). These fields correspond to the spin-flop field $H_{sf}$, which reaches a minimum value near $T_M$. This observation is consistent with the formula $H_{sf} = \lambda|M_{RE} - M_{TM}|$[36,37], where $M_{TM/RE}$ and $\lambda$ represent the magnetization of TM/RE sublattice and antiferromagnetic exchange interaction constant between TM and RE, respectively (see Supplementary Fig. S2). The magnetizations of Co and Gd after the spin-flop transition are oriented in the out-of-plane direction due to the perpendicular magnetic anisotropy and the gradual increase of AHE resistance indicates an inhomogeneity of the amorphous GdCo alloy film. The polarity reversal of the hysteresis loops and the spin-flop transitions were also confirmed by the AHE hysteresis loops under the out-of-plane magnetic field, as shown in Supplementary Fig. S1. Summarizing these results, the spin-flop state can be obtained above the external magnetic field of 8 T or higher in the temperature range between 220 and 290 K.

Next, we studied the response of the Néel vector to the induction of SOT under the spin-flop state of the GdCo alloy layer. In the experiment, the external magnetic field sufficient to induce spin-flop states was first applied in the in-plane direction, parallel to the spin current polarization, at the temperature near $T_M$. A pulsed current with 10-μs-duration was then applied to excite the Néel vector rotation. This pulsed current induces the SOT from the adjacent Pt and Ta layers into the GdCo alloy layer. Finally, the AHE resistance was measured by applying a direct current of 1 mA, which is small enough not to excite the rotation. Here, we emphasize that in the spin-flop states, the Néel vector is kept almost collinear with the Co magnetization and both are in the out-of-plane direction in the equilibrium. The sign change of the AHE resistivity indicates that the 180-degree rotation of the Co magnetization and the same rotation of the Néel vector, which allows us to argue the SOT-induced rotation of the Néel vector as detailed below.

Figure 3 shows the AHE resistance $R_{xy}$ measurements after applying the pulsed current for various system temperatures and current amplitudes with the magnetic field of 8 T or 9 T applied. At 220 K, the $R_{xy}$ signals remain stable regardless of the applied current because the magnetic field is insufficient for all the spins to undergo a spin-flop transition or the applied current does not reach the threshold required for the excitation of Néel vector rotation. On the other hand, the changes in $R_{xy}$ signal corresponding to the reversal of the Néel vector are observed between 230 and 260 K. This stochastic binary switching is most pronounced at higher currents and at the temperature near $T_M$. This can be understood as follows. When the pulsed current exceeds a threshold of the potential, such as the magnetic anisotropy, the Néel vector rotates during the pulsed current application and eventually relaxes along the out-of-plane direction, pointing randomly either upward or downward. Therefore, the observations of the stochastic binary switching imply the Néel vector ration driven by the SOT. It should be noted that at 260 K, the $R_{xy}$ signal occasionally takes intermediate resistance values. This is because the net magnetization is entirely compensated at 260 K and the magnetization can point in directions other than the out-of-plane direction[38,39]. In fact, the intermediate resistance values were observed only at the vicinity of $T_M$ during the flipping of the Néel vector, as shown in Supplementary Fig. S1(b).

No significant switching above 270 K is likely due to the temperature rise caused by the Joule heating during the pulse application. As shown in Supplementary Fig. S4, the temperature rises to around 300 K at the pulse amplitude of 20 mA, where the spin-flop state is released, and the Néel vector no longer rotates. In addition, the Néel vector rotation induced by SOT cannot be observed when the magnetic field is too small to induce the spin-flop state (see Supplementary Fig. S3).

To confirm the rotation frequency of the Néel vector rotation, a homodyne detection through the AHE was employed[40,41]. In this measurement, a Hall device fabricated from films cut from the same wafer as in the above experiments was used. Schematic of the measurement setup for the homodyne detection is shown in Fig. 4(a). We applied a bias current to excite the Néel vector rotation and a microwave current to detect the rotation frequency through a bias tee. Amplitude of the bias current was set to be 20 mA, which is enough to excite the Néel vector rotation at the temperature near $T_M$. To enhance the signal-to-noise ratio and to reduce Joule heating, the bias current was applied as a square waveform with a frequency of 50 kHz, and rectified Hall voltage was measured using a lock-in amplifier. To eliminate the non-resonant background signal, the resulting spectra were obtained by subtracting the data measured under 0 T from that under 8 T.

Figure 4(b) shows the rectified Hall voltage as a function of microwave frequency for various microwave power at 240 K, 250 K and 260 K, respectively. For the results of 240 and 250 K, distinct peaks appear at around 1-2 GHz. The peak amplitude increases with increasing the microwave power, which indicates that the resonance peak originates from the rectified signal of Néel vector rotation. The rotation frequency of the Néel vector is determined by the magnitude of the applied SOT and thus depends on the current value. The rotation frequencies observed in the experiments are in good agreement with the simulation results (see Supplementary S5). We also find that the peak frequency shifts toward higher frequencies as the microwave power increases. This shift could be attributed to an additional SOT from the microwave current and Joule heating at higher power. When the rotation is disturbed and momentarily stops due to falling below the threshold current, the center of peaks shifts to a higher frequency. Joule heating raises the device temperature closer to $T_M$ and consequently reduces the threshold current required to excite the Néel vector rotation. These factors explain the peak shift toward higher frequencies with increasing microwave power, however, further experiments are required to elucidate the mechanism in more detail. At 260 K, the peak disappeared above 12.5 dBm. This result corresponds to the fact that the stochastic binary switching of the anomalous Hall resistance disappeared above 270 K due to $H_{\text{ext}} < H_{\text{sf}}$. In addition, no peaks were observed in the Hall voltage measurement under conditions of smaller bias current (5 mA), lower microwave power (0 dBm), smaller external magnetic field (0.1 T), or lower temperature (10 K), as shown in Supplementary Fig. S6. From these observations, we confirm that the observed peaks result from the coupling between the Néel vector rotation and the applied microwave currents, and the rotation frequency of the Néel vector rotation is in the range of around 1-2 GHz.

In summary, we have demonstrated the excitation of the Néel vector rotation by SOT in spin-flopped ferrimagnetic GdCo alloy thin films. We revealed that the threshold current required for the Néel vector rotation depends on temperature, as evidenced by the stochastic binary changes in the Hall resistance by the application of a 10-$\mu$s pulsed current. Furthermore, we confirmed the rotation frequency of the Néel vector in around 1-2 GHz from the homodyne detection measurements, where the coupling between the Néel vector rotation and the microwave currents results in the rectified Hall voltage. In addition, the frequencies and threshold current values obtained in the experiments show a good agreement with simulated results. Our study provides a crucial advancement toward realizing spin superfluidity for low-dissipative spin transport.

## Acknowledgments

We acknowledge valuable discussion with Se Kwon Kim. This work was partially supported

by JSPS KAKENHI (JP20H05665, JP21K18145, JP24H00007 JP22H01936, JP24H02233), MEXT Initiative to Establish Next-generation Novel Integrated Circuits Centers (X-NICS) Grant Number JPJ011438, and Collaborative Research Program of the Institute for Chemical Research, Kyoto University.

**Figure Captions**

**Fig. 1.** The blue and orange arrows represent the local magnetic moments of two sublattices in ferrimagnet. These magnetic moments are fixed in a pseudo-magnetic easy-plane due to the spin-flop transition under the conditions in which the external magnetic field $H_{\text{ext}}$ is larger than the spin-flop field $H_{\text{sf}}$. The injected spin angular momentum $\sigma$ denoted by the red sphere with arrow causes the local magnetic moments to tilt slightly out of the easy-plane. This tilt generates an internal antiferromagnetic exchange field $H_{\text{ex}}$, which exerts an exchange torque $T_{\text{ex}}$ to rotate the Néel vector within the magnetic easy-plane.

**Fig. 2.** (a) Schematic illustration of the device and measurement setup. The directions of the electric current and the magnetic field are $x$- and $y$-direction, respectively. (b) The hysteresis loops of AHE resistance $R_{xy}$ from 220 K to 290 K when 1 mA was applied. The arrows indicate the spin-flop field at each temperature. The spin-flop field was determined by the peak for 250 K and 270 K, and by inflection points for other temperatures. The spin-flop field of 260 K cannot be determined because the value of the AHE resistance is almost constant.

**Fig. 3.** The AHE resistance $R_{xy}$ measured after 10-$\mu$s pulsed currents of various current amplitude were applied. The applied external magnetic field along $y$-direction and the system temperature are denoted at the top of each graph.

**Fig. 4.** (a) Schematic illustration of the measurement setup for homodyne detection through the anomalous Hall effect. (b) The rectified Hall voltage as a function of microwave frequency for various microwave power. The applied external magnetic field along the $y$-direction and the system temperature are denoted at the top of each graph. Circles indicate the resonance peaks attributed to the Néel vector rotation.

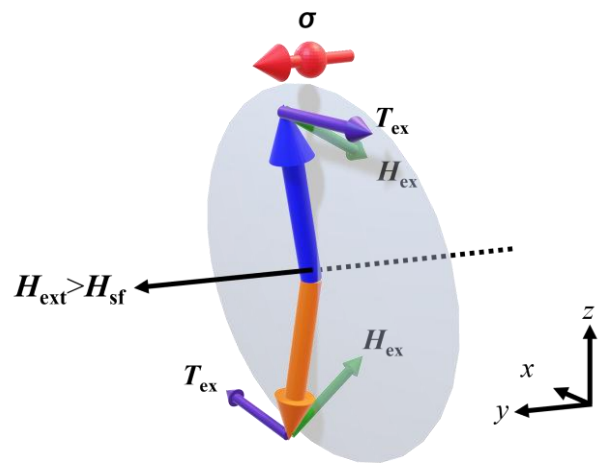

Fig. 1

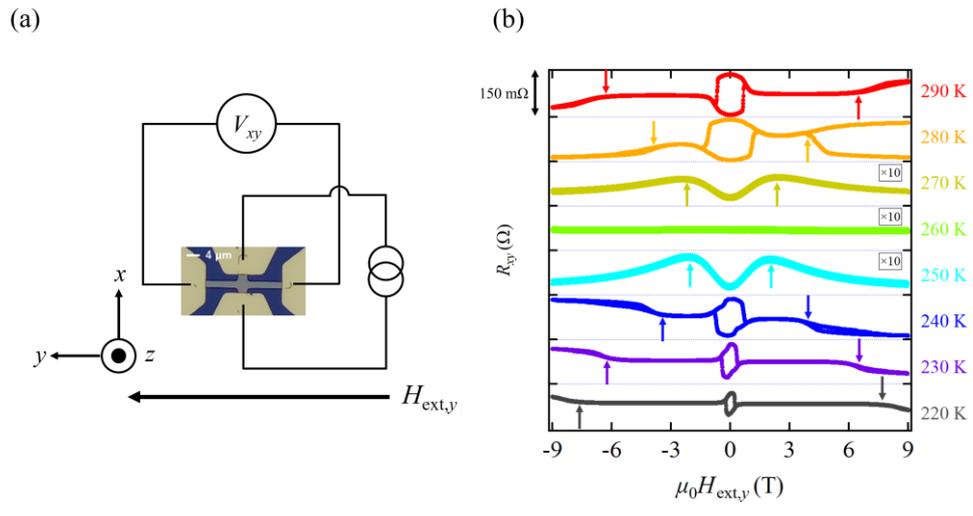

Fig.2.

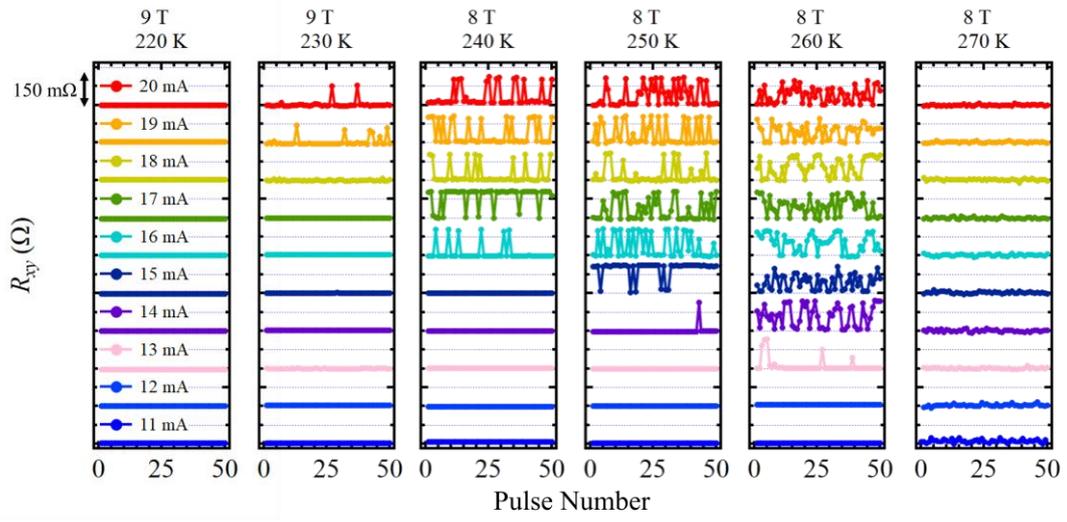

Fig. 3.

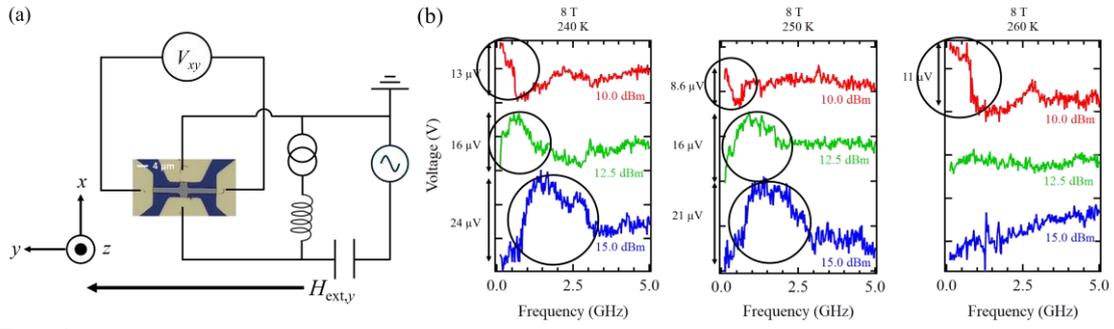

Fig. 4.

Supporting Information:

# Néel vector rotation driven by spin-orbit torque in amorphous ferrimagnetic GdCo

Tetsuma Mandokoro[1], Yoichi Shiota[1,2*], Tomoya Ito[1], Hiroki Matsumoto[1], Hideki Narita[1], Ryusuke Hisatomi[1,2], Shutaro Karube[1,2], and Teruo Ono[1,2*]

[1]*Institute for Chemical Research, Kyoto University, Uji, Kyoto 611-0011, Japan*

[2]*Center for Spintronics Research Network, Kyoto University, Uji, Kyoto 611-0011, Japan*

Authors to whom correspondence should be addressed: Yoichi Shiota, shiota-y@scl.kyoto-u.ac.jp; Teruo Ono, ono@scl.kyoto-u.ac.jp

## S1 Temperature dependence of AHE hysteresis loops under out-of-plane magnetic field

Figure S1(a) shows anomalous Hall effect (AHE) resistance $R_{xy}$ as a function of out-of-plane magnetic field at various temperatures denoted in the figure and Fig. S1(b) at the temperatures near $T_M$. The square hysteresis loops are clearly observed, indicating the perpendicular magnetic easy-axis in the temperature range from 230 K to 290 K. Notably, the jump in hysteresis is continuous at 265 K, a temperature near $T_M$. This result is consistent with the intermediate resistance values shown at 260 K in Fig. 3.

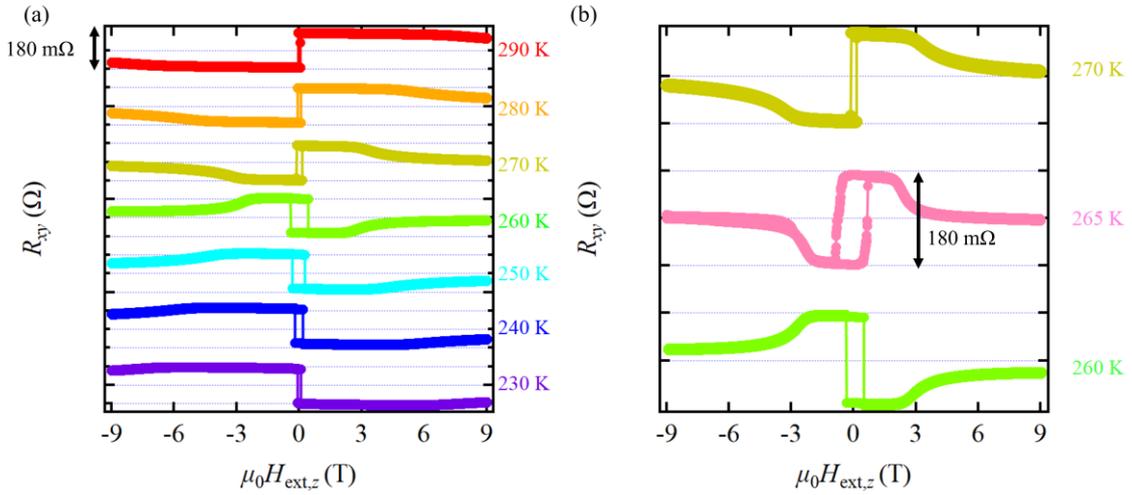

Fig. S1 $R_{xy}$-$H$ loops of Ta/GdCo/Pt under out-of-plane magnetic field in the temperature range from 230 K to 290 K.

**S2 Temperature dependence of spin-flop field**

Figure S2 shows the temperature dependence of the spin-flop field $H_{sf}$ under the in-plane magnetic field obtained from the results in Fig. 2(b) of the main text. The values of $\mu_0 H_{sf}$ was determined by the peak for 250 K and 270 K, and by inflection points for 220 K to 240 K, 280 K, and 290 K. The spin-flop field of 260 K cannot be determined because the value of $R_{xy}$ is almost constant. It was confirmed that the spin-flop field decreased as the temperature approached $T_M$.

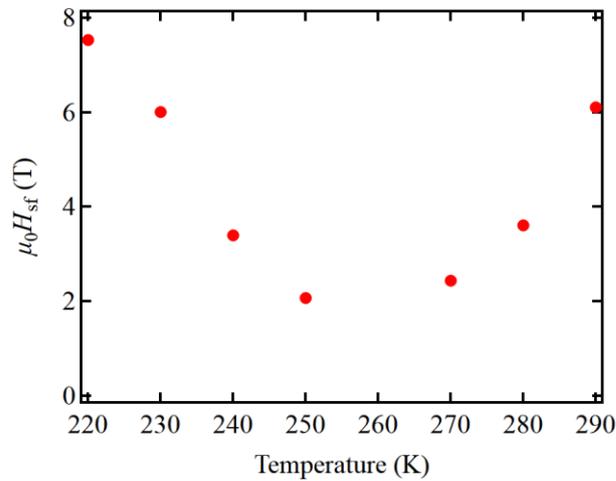

Fig. S2 Spin-flop field as a function of temperature obtained from the results in Fig. 2(b) of the main text.

**S3 Control experiments of the stochastic binary switching**

Figure S3 shows $R_{xy}$ as a function of pulsed current amplitude under the in-plane magnetic field ($y$-direction) of 0.1 T from 230 K to 260 K. No switching of the Néel vector was observed, because the magnetic field of 0.1 T is insufficient to induce the spin-flop transition.

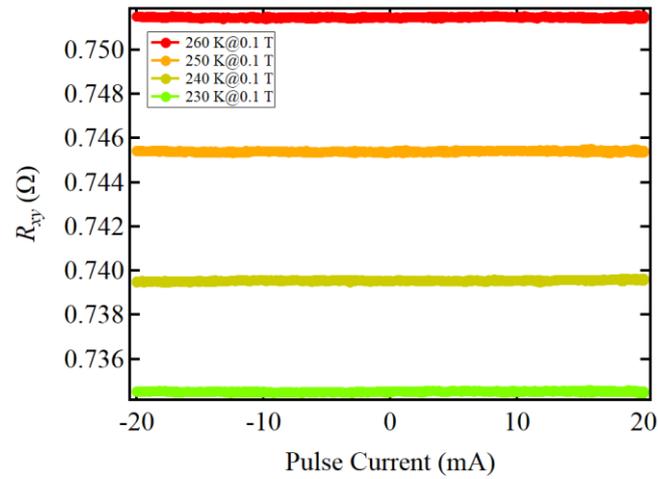

Fig. S3 AHE resistance $R_{xy}$ as a function of the current amplitude of 10-$\mu$s pulsed current when applied magnetic field is insufficient to the spin-flop transition.

**S4 Estimation of the temperature rise due to the Joule heating effect**

Figure S4 shows resistance of a wire $R_{xx}$ as a function of the temperature or the current to estimate the temperature rise when the direct current is applied at the system temperature from 220 K to 270 K. Since there exists the Joule heating effect, $R_{xx}$ exhibits a quadratic dependence on the current $I$. The solid lines are the best fit with $R_{xx}(I) = \sigma_I I^2 + \text{(constant)}$, where $\sigma_I$ is a quadratic constant for the electric current. The red circles represent the resistance $R_{xx}$ as a function of temperature $T$. The dashed line is the best fit with $R_{xx}(T) = \sigma_T T + \text{(constant)}$, where $\sigma_T$ is a proportionality constant for the temperature. In our device used for the stochastic binary switching experiment, $\sigma_T$ is estimated to be 0.0609 Ω K$^{-1}$. Combining these two measurements, we estimate the temperature rise as $\Delta T = (\sigma_I/\sigma_T)I^2$. The estimated $\sigma_I$ and the device temperature $T_{\text{device}}$ under the electric current from 10 to 20 mA at each system temperature is summarized in Table S1.

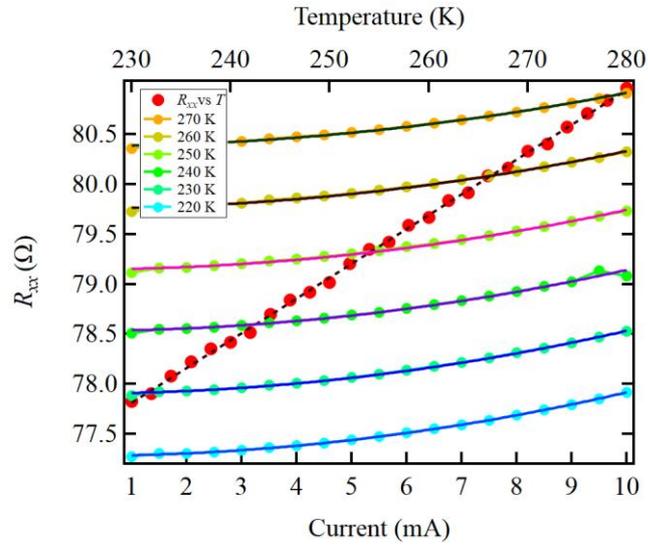

Fig. S4 Temperature dependence of $R_{xx}$ (red circles) and current value dependence of $R_{xx}$ at each system temperature.

Table S1. Estimation of temperature rise at each system temperature and current value

|  | 220 K | 230 K | 240 K | 250 K | 260 K | 270 K |
|---|---|---|---|---|---|---|
| $\sigma_I$ [Ω mA$^{-2}$] | 0.00637 | 0.00630 | 0.00609 | 0.00595 | 0.00573 | 0.00536 |
| $T_{\text{device}}$(10 mA) [K] | 232 | 242 | 251 | 261 | 271 | 280 |
| $T_{\text{device}}$(12 mA) [K] | 236 | 246 | 255 | 265 | 274 | 283 |
| $T_{\text{device}}$(14 mA) [K] | 241 | 251 | 261 | 270 | 279 | 288 |
| $T_{\text{device}}$(16 mA) [K] | 248 | 257 | 266 | 276 | 284 | 293 |
| $T_{\text{device}}$(18 mA) [K] | 254 | 264 | 273 | 282 | 291 | 299 |
| $T_{\text{device}}$(20 mA) [K] | 262 | 272 | 280 | 289 | 298 | 305 |

**S5 Simulation of Néel vector rotation driven by spin-orbit torque**

To confirm the validity of the Néel vector rotation observed in the experiment, we carried out the atomic spin simulations using "VAMPIRE" software[1,2]. We set the parameters in the simulation based on the experimental conditions at 240 K, as summarized in Table S2. The uniaxial magnetic anisotropy in the out-of-plane direction was experimentally determined to be $6.01 \times 10^3$ J/m$^3$, from the anomalous Hall resistance measured at 240 K when the magnetic field was swept in the in-plane direction[3]. The magnitude of magnetization for each sublattice at 240 K was obtained from the fitting of the temperature dependence of the magnetization[4]. Other parameters were used from literature values[5-11]. In the simulation setup, spin current was injected into a 3 nm-thick amorphous ferrimagnetic GdCo thin film from the top and bottom layers in an equal ratio. The in-plane direction of the unit cell was modeled under periodic boundary conditions. Due to the limitations of the simulation program, the decay of spin current was considered by dividing the thin film into 49 segments along the direction perpendicular to the surface. For simplicity, the spin torque was assumed to be only damping-like torque, and thermal fluctuations were neglected by setting the temperature to 0 K.

First, the system was initialized with random orientation of atomic magnetic moments and then relaxed under an external magnetic field of 8 T. The occurrence of spin-flop was confirmed under these conditions. Simultaneously, an electric current was applied to the system to induce spin-orbit torque, allowing for the computation of the Néel vector dynamics. Finally, the rotation frequencies of the Néel vector were evaluated for various current.

The dependence of the rotation frequency on the dc current is shown in Fig. S5(a). The red, green, and blue dots show the results for damping constants of 0.02, 0.06, and 0.1, respectively. The dependence of the rotation frequency on the damping constants when 20 mA is applied is presented in Fig. S5(b). At a current value of 20 mA, corresponding to the experimental conditions, the calculated rotation frequencies are 1.78 GHz, 2.97 GHz, and 9.05 GHz for the damping constants of 0.1, 0.06, and 0.02, respectively. The experimentally obtained rotation frequency in the 1-2 GHz range closely matches the calculated values for damping constant between 0.05 to 0.1. These damping constant values are reasonable for amorphous ferrimagnetic GdCo compared to those reported in previous studies[49,53].

Additionally, the threshold current for all damping constants in the simulation, 14 mA, closely matches the threshold current shown in Fig. 3.

Table S2. Parameters used in simulations.

| Parameter | Value (Co/Gd) |
|---|---|
| Atomic composition, [%] | 72 / 28 |
| Effective spin Hall angle, $\theta_{SH}$ [3] | 0.21 / 0.03 |
| Magnetization, $[\mu_B]$ [4] | 1.85 / 4.85 |
| Gyromagnetic ratio, $\gamma$ [rad Hz/T] [5] | $1.86\times10^{11}$ / $1.78\times10^{11}$ |
| Intra-lattice exchange energy, [J] [6] | $111\times10^{23}$ / $2.8\times10^{23}$ |
| Inter-lattice exchange energy, [J] [6] | $-24\times10^{23}$ |
| Spin diffusion length [nm] [7] | 5 |
| Gilbert damping constant, $\alpha$ [6,8-11] | 0.01-0.10 |

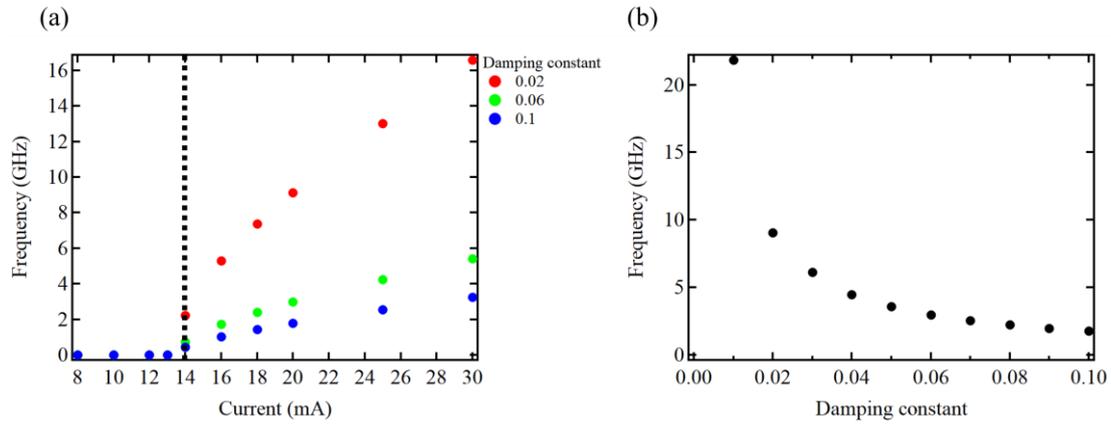

Fig. S5 (a) Calculated rotation frequencies of Néel vector as a function of current amplitude for various damping constants. (b) Calculated rotation frequencies of Néel vector as a function of damping constant when 20 mA is applied.

**S6 Control experiments of the homodyne measurements**

Figure S6 shows the rectified Hall voltage as a function of microwave frequency for various conditions as referred in legend. The black line shows the result adapted from Fig. 4(b) in the main text, where the Néel vector rotation occurs. However, no clear peak is observed when the electric current does not reach the threshold required for the excitation of Néel vector rotation (red), when the microwave power is small (blue), and when there is no spin-flop transition due to insufficient magnetic field (green) and the temperature far from $T_\mathrm{M}$ (purple).

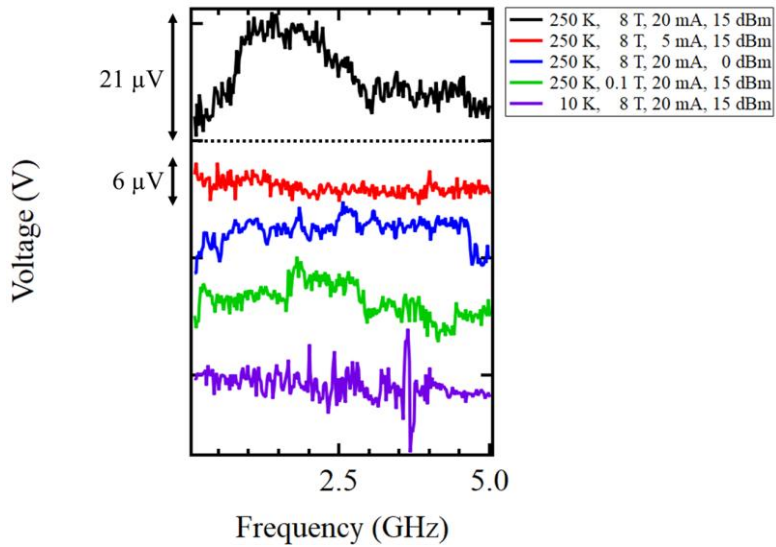

Fig. S6 Control experiments examining homodyne detection measurements under various conditions.

## Supporting References